\documentclass[letter]{aa}  

\usepackage{graphicx}
\usepackage{txfonts}
\usepackage{lipsum}
\usepackage{float}
\usepackage{sidecap}
\usepackage{subcaption}         
\usepackage{lscape}             
\usepackage{placeins}           
                                
\newcommand{\dcop}{$\rm DCO^+$\xspace}

\newcommand{\nndp}{$\rm N_2D^+$\xspace}

\newcommand{\nnhp}{$\rm N_2H^+$\xspace}
\newcommand{\kms}{$\rm km \, s^{-1}$\xspace}

\begin{document}

   \title{Corona Australis 151: an extremely young protostar}


%
%
%

\author{E. Redaelli\inst{1} \and S. Spezzano\inst{2}   \and P. Caselli\inst{2} \and J. Harju \inst{3,2} \and D. Arzoumanian \inst{4,5} \and O. Sipil\"a\inst{2} \and A. Belloche\inst{6} \and F. Wyrowski\inst{6} \and J. E. Pineda \inst{2} \and S. Jensen\inst{2} \and Y. Misugi\inst{7,8} \and G. Majerczyk\inst{9,1}}  

\institute{European Southern Observatory, Karl-Schwarzschild-Strasse 2, 85748 Garching, Germany \and Max-Planck-Institut f\"ur Extraterrestrische Physik, Giessenbachstrasse 1, 85748 Garching, Germany \and Department of Physics, P.O. Box 64, FI-00014, University of Helsinki, Finland  \and
Institute for Advanced Study, Kyushu University, Japan 
\and Department of Earth and Planetary Sciences, Faculty of Science, Kyushu University, Nishi-ku, Fukuoka 819-0395, Japan 
\and
    Max-Planck-Institut f\"ur Radioastronomie, Auf dem H\"ugel, 69, 53121 Bonn, Germany
  \and
  Faculty of Science and Engineering, Kyushu Sangyo University, 2-3-1 Matsukadai, Fukuoka 813-8503, Japan 
  \and National Astronomical Observatory of Japan, Osawa 2-21-1, Mitaka, Tokyo 181-8588, Japan
  \and     
 Department of Physics, University of Rome Sapienza, P.le A. Moro 5, 00185, Rome, Italy
}

   \date{XXX}

 
  \abstract
   {Prestellar cores are the birthplace of stars and planetary systems, but they are short-lived objects, since the initial stages of dense core evolution, collapse, and the formation of a protostellar seed are fast. In an effort to build a catalogue of bona-fide prestellar cores in the Solar neighbourhood, we used APEX observations to identify dynamically evolved cores among dense cores observed with \textit{Herschel}. One of them, Corona Australis 151, stood out because of its centrally peaked structure, with densities above $10^7\, \rm cm^{-3}$ in the central $500-1000\,$au, and high deuteration levels ($\rm N_2D^+/N_2H^+ \sim 0.5$){, suggestive of an evolved, prestellar stage}.}
   {Corona Australis 151 appears to be an evolved prestellar core, but the presence of broad wings in some of the detected lines and a tentative 70$\,\mu$m detection with \textit{Herschel} partially challenges its prestellar stage. We intend to assess its evolutionary stage.} 
   {We analyse new ALMA data of the continuum emission and of several line tracers (including typical outflow tracers) at a resolution of $\sim 150-200\, \rm au$, tracing the envelope. }
   {We unveil the presence of a compact and young outflow (projected dynamical age: $\sim 500 \, \rm yr$), traced by SiO, CO, and $\rm H_2CO$ emission. The continuum emission traces an envelope structure of size $\sim 1000\,$au and mass $0.33 \, M_\odot$ (assuming $T_\mathrm{dust}=20\,$K). }
   {Corona Australis 151 is an extremely young protostar, possibly one of the youngest known in the Solar neighbourhood, and hence a new laboratory to study the chemical and dynamical evolution at the dawn of star formation.}

   \keywords{ISM: clouds --- ISM: molecules --- Stars: formation --- ISM: individual objects: Corona Australis 151
               }

   \maketitle
\nolinenumbers

\section{Introduction}
Prestellar cores are {gravitationally bound starless cores} on the verge of gravitational collapse. They contain all the ingredients to form stars and planetary systems, so their study provides critical information on the initial conditions of the assembly of stellar systems like our own. They differ from the more general class of starless cores (dense cores not hosting young stellar objects or YSOs) due to their large central density $n \rm (H_2) > 10^5 \, \rm cm^{-3}$ and centrally concentrated density distributions \citep{Crapsi05, keto08}.  Prestellar cores are rare, as their lifetimes are short ($\sim 50,000 \,$yr, see e.g. \citealt{Tassis07}). L1544, embedded in the Taurus molecular cloud, represents a prototypical example of a prestellar core, and it has been the target of several observational studies \citep{Caselli02b, Spezzano16, Spezzano17, Redaelli19a, Redaelli21b, Caselli22, Bianchi23}. In an effort to build a large sample of prestellar cores in the Solar neighbourhood, in \cite{Caselli25} we compiled a catalogue of 40 dense cores extracted from the \textit{Herschel} Gould Belt Survey Archive (HGBS; \citealt{Andre10}), by selecting high-density objects ($n_\mathrm{central}> 3 \times 10^5 \, \rm cm^{-3}$ in the central $20''$) with no known associated protostellar candidates. The project used high-frequency ($0.7-1.3\,$mm) observations from the Atacama Pathfinder EXperiment (APEX; \citealt{Gusten06}) to characterise the high-J emission of the well-known high-density tracers \nnhp and \nndp (critical densities $n_\mathrm{crit} \gtrsim 10^6 \, \rm cm^{-3}$). Only three targets were detected in the high-frequency transitions \nnhp $5-4$ and \nndp $6-5$, including Corona Australis 151 (hereafter CrA 151).  

This source sits within region CrA-E described by \cite{Bresnahan18}, in the northern filament of the cloud, at about 5$\,$pc east of the well-known young Coronet protocluster. The Gaia-based distance is $d = 150\,$pc \citep{Galli20}. No YSO is known in its proximity (cf. \citealt{Esplin22}). However, \cite{Bresnahan18} detected a point source at the $4\sigma$ level in the \textit{Herschel}/PACS 70$\, \mu$m map. This could potentially indicate the presence of a very low-luminosity object (VeLLO, \citealt{diFrancesco07, Tomida10}; cf. \citealt{Bourke06} in L1521F), either a very young low-mass protostar or a first hydrostatic core (FHSC, \citealt{Larson69}). The FHSC stage is a short-lived ($\sim 10^4\,$yr) phase preceding the formation of the protostar (see simulations from \citealt{Commercon12, Tomida13}). Simulations show that FHSCs can launch poorly collimated and slow (a few \kms) outflows (see e.g. \citealt{Machida08}), {which however can play an important role in the initial evolution of the systems, removing angular momentum efficiently \citep{Machida14}. Observationally, the search for FHSC candidates has been quite difficult, with several works searching for them \citep{Pineda11,Pezzuto12,Busch20}}.
\par
In \cite{Redaelli25}, we modelled \textit{Herschel} continuum maps and APEX line data, showing that \textit{i)} CrA 151 is dense ($n ({\rm H_2})>10^7 \, \rm cm^{-3}$ at scales of $\sim 1000\, $au); \textit{ii)} it presents a very high deuteration level of \nndp/\nnhp$\sim 50 $\%, with several other deuterated species detected, such as $\rm HDO$, $\rm CHD_2OH$, $\rm D_2CO$, and even $\rm ND_3$; \textit{iii)} the \nnhp and \nndp lines present narrow linewidths, indicative of low levels of turbulence and low temperatures. These results are consistent with an evolved prestellar core, since deuterated molecules are very abundant under high-density, low-temperature conditions \citep{Ceccarelli14}. \par
A few details, however, appear to contradict the aforementioned scenario. In particular, several sulphur-bearing species present in the APEX frequency coverage show line profiles with extended, high-velocity (a few \kms with respect to the source $V_\mathrm{lsr}$) wings. Furthermore, the modelling of the \nnhp isotopologues unveiled no depletion of these species even in the central part of the core, unlike other similarly evolved prestellar cores (e.g. L1544, \citealt{Redaelli19a}). The question of whether CrA 151 is an evolved prestellar core, a FHSC, or an extremely young YSO remained unanswered. In this Letter, we report new observations taken with the Atacama Large Millimeter and sub-millimeter Array (ALMA) at $\sim 200 \,$au resolution that show the presence of a young protostar at the centre of the core.
   \begin{figure*}[!h]
   \centering
   \includegraphics[width=.95\hsize]{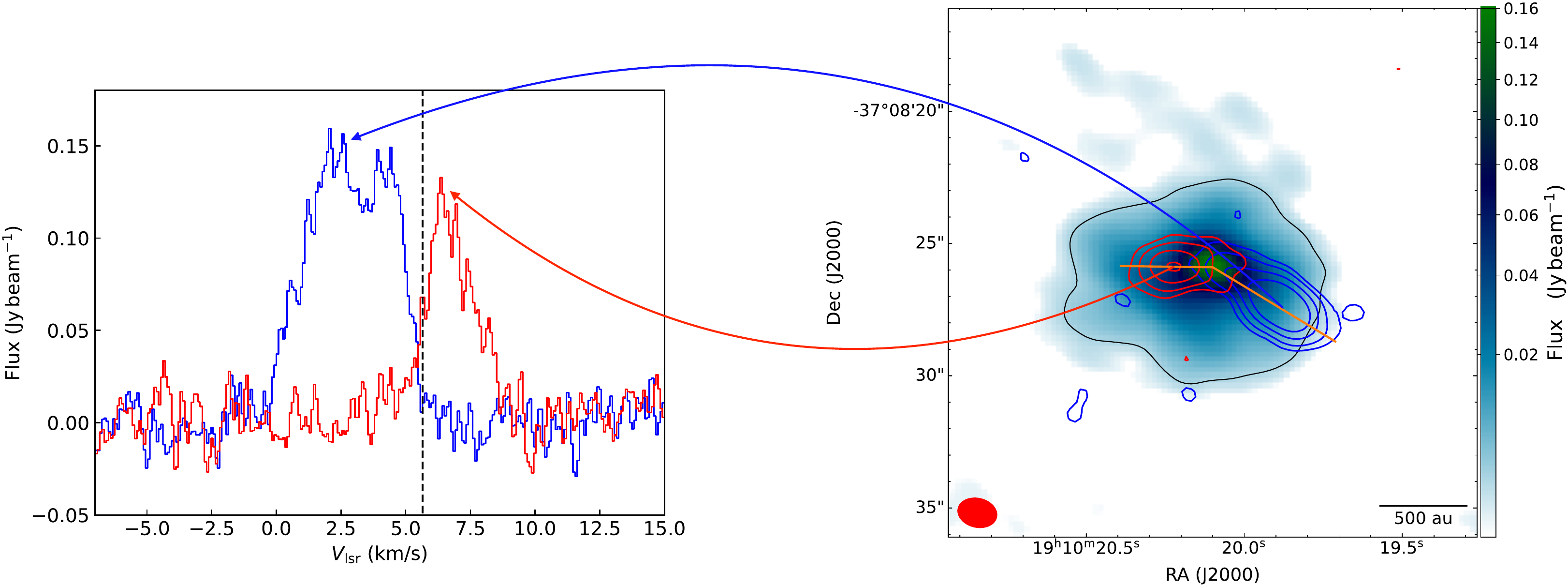} 
            \caption{\textit{Right panel:} ALMA Band 6 emission towards CrA 151. The {black} contour shows the $\rm S/N>5$ region. The red and blue contours show the integrated emission of the redshifted and blue-shifted components of the SiO $5-4$ emission, computed in the ranges  $[5.7;10.5]$\kms and $[-1;5.6]$\kms, respectively. The contours represent the $[5,10,20,35]\sigma$ levels. The beam size (continuum) and scalebar are shown in the bottom-left and -right corners. {The orange line shows the cut used to produce the PV plot in Fig.~\ref{fig:PV_plot}.} \textit{Left panel:} the histograms show two SiO spectra, extracted at the peaks of the blue and red lobes. The vertical dashed line shows the source $V_\mathrm{lsr}=5.65\,$\kms \citep{Redaelli25}.}
         \label{fig:cont}
   \end{figure*}
\section{Observations}
ALMA Band 6 observations towards CrA 151 ($\rm RA(J2000) = 19^h10^m20^s.17$, $\rm Dec(J2000) = -37^\circ 08^m 27^s.0$, ICRS frame) were performed, using both the 12m-array (in configuration C43-2) and the 7m-array, in November 2024 and May 2025, as part of project ID 2024.1.00605.S (PI: D. Arzoumanian). The baselines range from $8.9\,$m to $314\,$m {and the maximum recoverable scale is $\sim$30$''$ (corresponding to $\sim$4500$\,$au)}. The bright quasar J1924-2914 was used as amplitude and bandpass calibrator, whilst J1937-3958 and J1925-3401 were used as phase calibrators for the 7m and 12m arrays, respectively. We performed a mosaic of 3 pointings with the 7m-array and 7 pointings with the 12m-array. The spectral setup consists of 5 spectral windows (SPWs): two high-resolution (channel width $61.5\,$kHz, or $0.08\,$\kms) SPWs are centred on the \dcop $3-2$ and $\rm SiO$ $5-4$ transitions; the remaining three SPWs were dedicated to continuum observations, with a bandwidth of $1.875\,$GHz and a coarse channel width of $1.3\,$\kms. These are centred at $219$, $230$, and $233\,$GHz.
\par
Details regarding the data reduction and imaging are summarised in Appendix~\ref{app:imaging}. The final synthesized beam sizes of the imaged data are $1''.49\times 1''.09 \;(\rm PA=76.9^\circ)$ (continuum), $1''.60\times 1''.19 \;(\rm PA=76.9^\circ)$ (SiO $5-4$), and $1''.63\times 1''.21 \;(\rm PA=78.9^\circ)$ (\dcop $3-2$). The continuum sensitivity is $0.6\, \rm mJy \, beam^{-1}$, whilst the sensitivity of the high-resolution spectral windows is $11-12\, \rm mJy \, beam^{-1}$. These values refer to non-primary-beam corrected data; throughout the paper, we generally refer to data without primary-beam correction, unless otherwise stated. \par

The continuum-dedicated SPWs cover also the CO $2-1$ line and the $\rm H_2CO$ $3_{0,3}-2_{0,2}$ line, at a coarse velocity resolution ($978\,$kHz, or $\approx 1.3\,$\kms). We imaged these transitions at a sensitivity of $3.4 \rm \, mJy\, beam^{-1}$ per channel.

\section{Analysis and Discussion}
{Figure~\ref{fig:cont} shows the continuum emission in colorscale, which consists of a fairly compact, centrally peaked structure of $\approx 1000\,$au in size, with a peak flux of $\approx 160 \, \rm mJy \, beam^{-1}$. The spectra of SiO $5-4$, extracted towards the east and southwest of the continuum peak, both present broad spectral shapes that are red-shifted and blue-shifted, respectively, with respect to the source local-standard-of-rest velocity ($V_\mathrm{lsr}=5.65\,$\kms based on \nnhp and \nndp data; see \citealt{Redaelli25}). These features clearly unveil the presence of an outflow in CrA 151.}

\subsection{The continuum emission\label{sec:cont}}
The mass of the CrA 151 envelope, traced by the dust continuum emission, is estimated using the standard equation:
   \begin{equation}
    \label{eq:Menv} 
M_{\rm env}= f \; \frac{d^2 \, S_{\rm 1.3mm}}{k_{\rm 1.3mm} B_{\rm 1.3mm}(T_\mathrm{dust})}.
   \end{equation}
Here, $S_{\rm 1.3mm} =663 \, \rm mJy$ is the total flux in the primary-beam corrected image, which we compute from pixels with $\rm S/N>5$ (the white contour in Fig.~\ref{fig:cont}), $k_{\rm 1.3mm}=0.9 \, \rm cm^{2} \, g^{-1}$ is the dust opacity \citep{Ossenkopf94}, $d$ the source's distance, $f = 100$ the {gas-to-dust} mass ratio, and $B_{\rm 1.3mm}(T_\mathrm{dust})$ the Planck function at the dust temperature $T_\mathrm{dust}$. We adopt $T_\mathrm{dust}=20\,$K, compatible with predictions from models of Class 0 envelopes \citep{Shirley02}. For comparison, the brightness temperature of the continuum peak is $T_\mathrm{b} = 14\,$K, which can be considered a lower limit to the dust temperature. With these values, we obtain $M_\mathrm{env}=(0.33
\pm 0.06)\rm \, M_\odot$, where we assume a conservative 20\% error to take into consideration the uncertainties on the flux, on $k_{\nu}$, and on $T_\mathrm{dust}$. This value is consistent with envelope masses of similarly young cores measured with interferometric observations \citep{Stephens18, Maureira20}. Given the equivalent radius of the envelope of $630 \, $au, this corresponds to an average density in the envelope of $4\times 10^7 \, \rm cm^{-3}$.
\subsection{The outflow properties}
To investigate the outflow properties, we compute the integrated intensities of the blue-shifted and red-shifted components only, and we show them in Fig.~\ref{fig:cont}. 
 The red lobe appears spatially concentrated, with an extension of $\approx 600 \, \rm au$ in the West-East direction in the plane of the sky. The blue lobe is instead brighter and more extended, stretching for $\approx 900 \,$au towards south-west and beyond the edges of the continuum emission. We can speculate that, if the continuum brightness traces the matter density distribution, the blue-shifted lobe bursts through a region of lower density, hence expanding more. We estimate the projected dynamic age of the outflow based on the length of the blue lobe ($L=900\,$au, based on the $5\sigma$ contour in Fig.~\ref{fig:cont}) and its velocity extension $\Delta V = 8.5\,$\kms, computed looking at the average SiO spectra in the same region (spectrum not shown here) and considering the positions in velocity where the flux reaches the zero level. With this data, we evaluate $t_\mathrm{dyn}^\mathrm{pr}= L/\Delta V = 500\,$yr. {This value is the projected quantity\footnote{See \cite{Dunham14} for a discussion on the inclination effects on outflow properties.}, which is related to the intrinsic one through the inclination of the system with respect to the line-of-sight ($\phi$):  $t_\mathrm{dyn} = t_\mathrm{dyn}^\mathrm{pr}/\tan(\phi)$. The presence of SiO emission arising from a currently ejected outflow suggests that the outflow has a velocity $\gtrsim 20\,$\kms, as this is the minimum velocity required to release Si from dust grains \citep[][and references therein]{Gusdorf08}. By contrast, the SiO abundance in quiescent gas is low \citep[$<10^{-12}$,][]{Ziurys89}. Given the observed velocity of $8.5 \,$\kms, we estimate $\phi \geq 65^{\circ}$. {The dynamical age corrected for this value is $t_\mathrm{dyn} \leq 250\,$yr}.
\par
Formaldehyde and CO also trace the outflow. The $\rm H_2CO$ line presents an arc-shaped extended emission in the eastern part of the source, which is visible in the central channels (see Fig.~\ref{fig:H2CO}). The feature, which is also partially visible in the CO data (cf. Fig.~\ref{fig:CO_chmap}), potentially resembles a streamer, similarly to the ones that are becoming frequently observed towards young stellar objects and discs. In particular, several have been found already in embedded Class 0 objects \citep{Pineda20, Thieme22, Murillo22, ValdiviaMena23, Taniguchi24}. Since the feature is visible only in two channels {(4.8 and 6.1$\,$\kms in Fig.~\ref{fig:H2CO})}, a detailed kinematic analysis is not possible at this stage, hampering a conclusive assessment of its nature{. It might also be part of the outflow}.
\subsection{The envelope in molecular emission}
The \dcop $3-2$ emission is well extended over the whole field-of-view, hence the importance of combining the interferometric data with the single-dish ones. In the following, we discuss and model the primary-beam corrected and combined data. The line shape is a single Gaussian in the external parts of the envelope, with narrow linewidths (see Appendix~\ref{app:dcop} for more details). In the central $1000-1500\,$au, however, a second, fainter velocity component is visible at higher velocities {($>6 \,$\kms, see Figs.~\ref{fig:dcop_channelmap} and \ref{fig:dcop_spectra})}. Assuming $V_\mathrm{lsr}=5.65\,$\kms, the flux dip does not appear in correspondence with the source's velocity (cf. Fig.~\ref{fig:dcop_spectra}), suggesting that the second feature is an additional velocity component, and not due to e.g. self-absorption. The weaker component, in fact, presents velocities in the range $5.9-6.2\,$\kms. \par
To model the \dcop spectra, we start by fitting a single Gaussian to all positions where $T_{peak}/rms > 10$. Next, we calculate residuals within the velocity range $[3.9-8.2]\,$\kms and perform a second fit with two Gaussians for pixels where residuals exceed $ 2\times rms$. We then create composite images for the centroid velocity (see Fig.~\ref{fig:dcop_vel}) and velocity dispersion (shown in Fig.~\ref{fig:dcop_sigma}), which combine the single Gaussian results when two components cannot be fitted and the blueshifted velocity component results elsewhere. \par
The line centroid shows a smooth gradient in the direction north-west towards south-east, with values going from $5.2\,$\kms to $5.8 \,$\kms. This direction is qualitatively perpendicular to that of the outflows, which is suggestive of core and envelope rotation. We measure a velocity gradient of $25\, $\kms$\, \rm pc^{-1}$, in agreement with similar measurements performed with interferometers in Class 0 envelopes \citep{Gaudel20}.
\par
The \dcop lines are narrow: the measured velocity dispersion is $\sigma_\mathrm{V} \leq 0.3 \,$\kms. The increase seen towards the centre is associated with the appearance of the second velocity component, which --when not properly resolved-- leads to line broadening. The median value is $\langle \sigma_\mathrm{V} \rangle = (0.18\pm 0.02) $\kms. For comparison, the sound speed at $20\,$K is $C_\mathrm{S} = 0.26\,$\kms. The thermal linewidth of \dcop at the same temperature is $\sigma_\mathrm{th} = 0.07\, $\kms. The envelope is quiescent with subsonic motions. 
   \begin{figure*}[h!]
   \centering
   \includegraphics[width=.95\hsize]{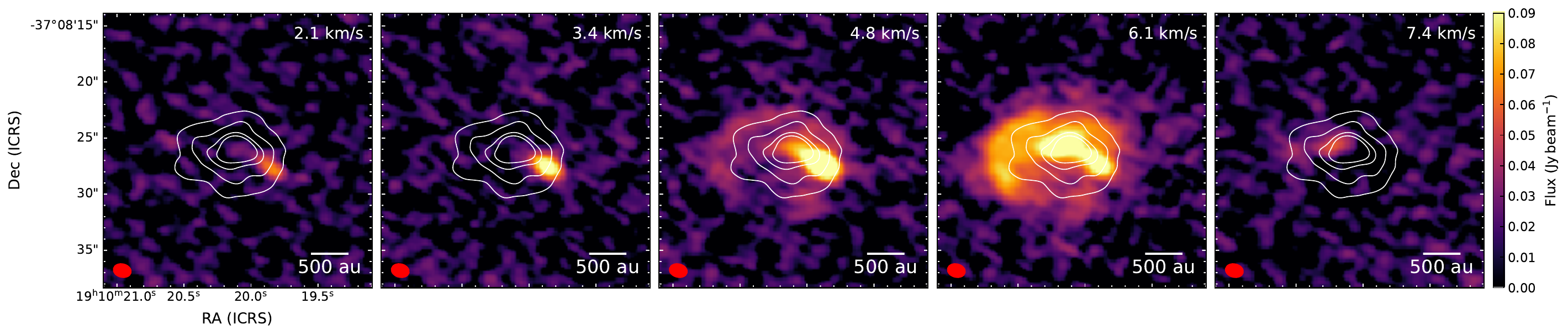}
      \caption{Channel maps from the $\rm H_2CO$ $3_{0,3}-2_{0,2}$ cube. The velocity of each panel is labelled in the top-right corner, whilst the beam size and the scalebar are shown in the bottom-left and right corners, respectively. The white contours show the continuum emission (levels: $[5,20,50,80]\sigma$). The first three channels show the bright blue outflow lobe, whilst the red lobe is visible mainly in the last two. The maps at $4.8$ and $6.1\,$\kms, however, present more extended emission, which appears in the shape of an arc-like feature to the east of the continuum envelope. }
         \label{fig:H2CO}
   \end{figure*}
      \begin{figure}
   \centering
   \includegraphics[width=\hsize]{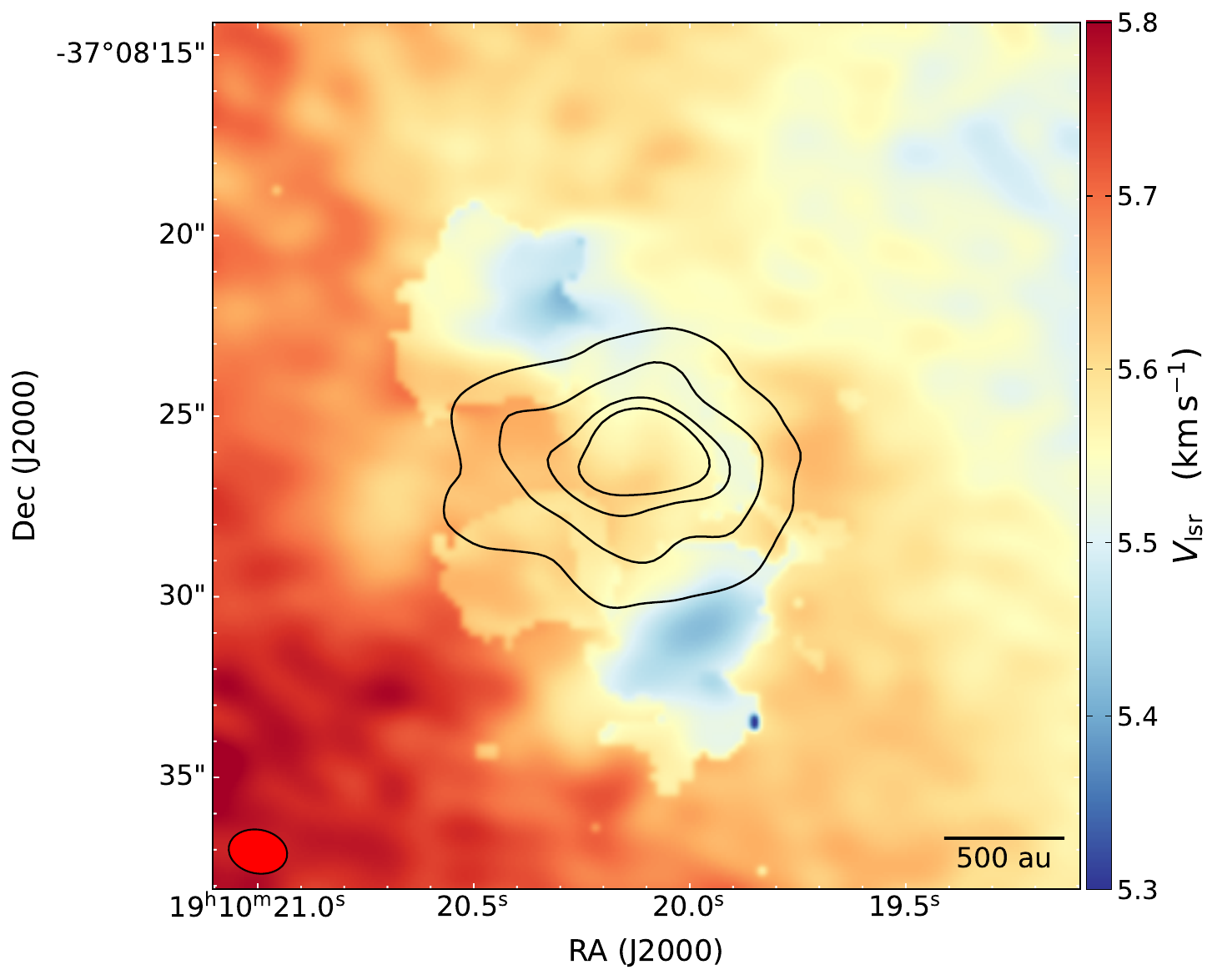}
      \caption{Centroid velocity obtained by fitting one or two Gaussian component(s) to the \dcop $3-2$ line, as described in the main text. The abrupt changes happen at the transition from the 1- to the 2-component fit.       }
         \label{fig:dcop_vel}
   \end{figure}

\section{Conclusions}
In this Letter, we addressed the evolutionary stage of the core CrA 151. The ALMA data unambiguously reveal the presence of a molecular outflow, seen in multiple species: SiO, formaldehyde, and CO. {The presence of a compact SiO outflow strongly favours a protostellar nature over an FHSC interpretation; under the standard assumption that SiO requires shock velocities of $\gtrsim 20\,$\kms, the observed flow is difficult to reconcile with FHSC models predicting only a few \kms.} The core bolometric temperature, evaluated from the \textit{Herschel} data, is $18.8 \, \rm K$. This is well below $70\, $K, which is considered the limit between Class 0 and Class I objects \citep{Chen95}.
\par
Several pieces of evidence support that the protostar at the centre of CrA 151 is extremely young. We {derived a projected dynamical age of 500$\,$yr for the outflow. Correcting for the inclination value obtained from the SiO data, we compute $t_\mathrm{dyn}\leq 250\,$yr.} {The \dcop emission traces a quiescent envelope with an ordered velocity gradient, indicating that the dense gas surrounding the protostellar object is still unperturbed by the protostellar activity, as supported by the high level of \nnhp deuteration seen in single-dish data \citep{Redaelli25}.} 
\par
CrA 151 appears to be an ideal new laboratory to study young protostars and their surroundings. High-angular-resolution ($\sim 10 \,$au) data will be necessary to detect and possibly model the disk properties and to constrain the mass of the central object. This analysis will provide, among others, the inclination of the system, which is essential to obtain reliable values for the outflow properties. Furthermore, the intermediate disk-envelope scales also deserve attention. The $\rm H_2CO$ and CO data show extended emission in the shape of an arc-like structure. Still, we lack the spectral resolution and large-scale flux information needed to analyse this feature kinematically. This work highlights the importance of the synergy between single-dish and interferometric observations to probe the early phases of star formation.
\begin{acknowledgements}
This paper makes use of the following ALMA data: ADS/JAO.ALMA\#2024.1.00605.S ALMA is a partnership of ESO (representing its member states), NSF (USA) and NINS (Japan), together with NRC (Canada), NSTC and ASIAA (Taiwan), and KASI (Republic of Korea), in cooperation with the Republic of Chile. The Joint ALMA Observatory is operated by ESO, AUI/NRAO and NAOJ.
\end{acknowledgements}

\bibliographystyle{aa}
\bibliography{Literature}

\begin{appendix}
\nolinenumbers
\twocolumn
\section{ALMA data reduction and imaging \label{app:imaging}}

\FloatBarrier
   \begin{figure}[!h]
   \centering
   \includegraphics[width=.4\textwidth]{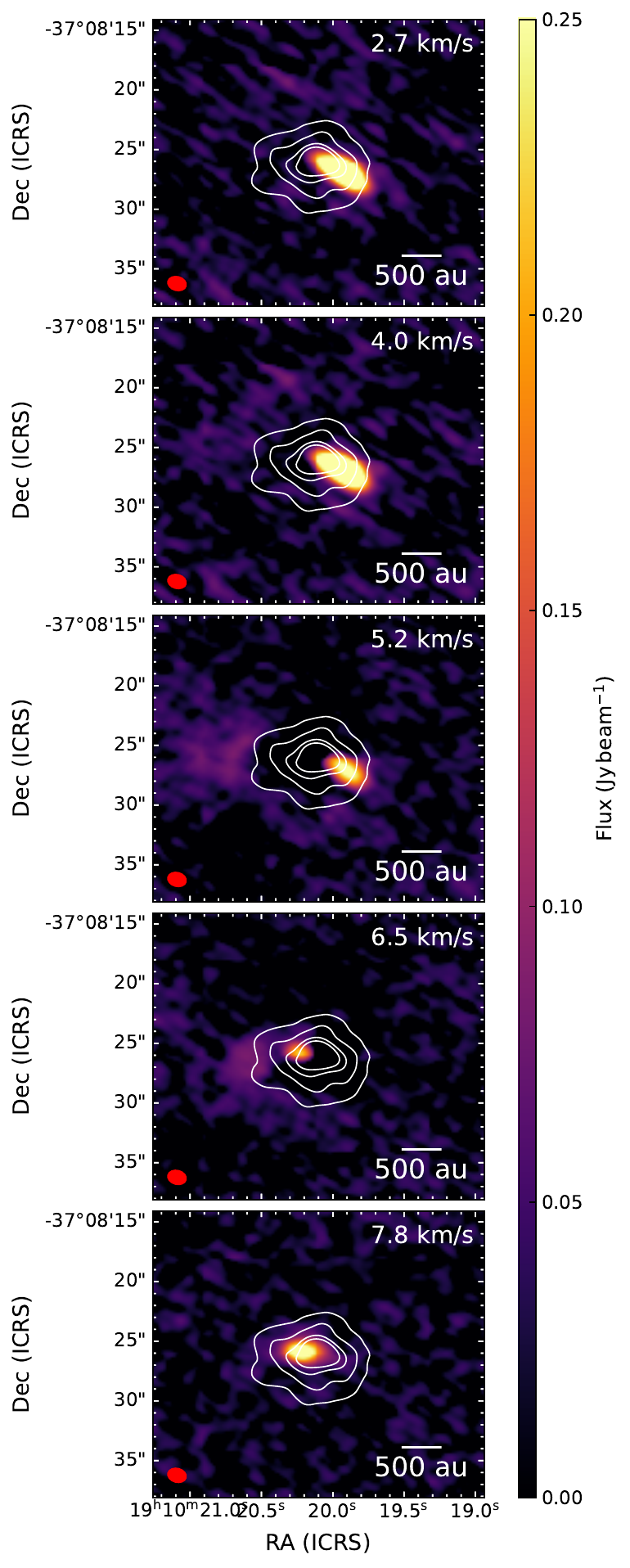}
      \caption{Same as Fig.~\ref{fig:H2CO}, but for the CO $2-1$ transition. The third and fourth panels show extended, fainter emission to the east of the continuum emission boundaries, similarly to the arc-like feature seen in formaldehyde. Especially around the source $V_\mathrm{lsr}$, several negative bowls are visible, due to the missing-flux problem, as a consequence of the lack of large-scale emission sensitivity.}
         \label{fig:CO_chmap}
   \end{figure}
The data were calibrated by the pipeline (version 2024.1.0.8). The rest frequencies of the molecular transitions are: \dcop $3-2$, 216112.5822$\,$MHz; SiO $5-4$, 217104.9190$\,$MHz; CO $2-1$, 230538.0$\,$MHz; and $\rm H_2CO$ $3_{0,3}-2_{0,2}$, 218222.1920$\,$MHz\footnote{All the rest frequencies are taken from the CDMS database, available at \url{https://cdms.astro.uni-koeln.de/}. The relevant laboratory publications are \cite{Winnewisser97, Brunken03, Caselli05, Muller13,Muller17}.}.The imaging was performed using \textsc{casa} version $6.7.3$. For the continuum emission and the SiO, $\rm H_2CO$, and CO lines, we cleaned the 7m- and 12m-data simultaneously using \textsc{tclean}. In general, we performed interactive cleaning to set the masks manually. We always used \textsc{weighting='briggs'} with \textsc{robust=0.5}. The \textsc{datacolumn='corrected'} was selected to use self-calibrated data when available. The pixel size was set to $0''.2$ (i.e. approximately 1/5 of the beam minor axis).
\par

For the continuum, we used \textsc{deconvolver='hogbom'}. We took the line-free channels delivered by the pipeline in the weblogs. For the SiO, $\rm H_2CO$, and CO transitions we used \textsc{deconvolver='multiscale'},  with scales = [0,5,15,30]. In general, we always prefer multiscale when the emission is extended over several beams. However, in the case of the continuum (and the \dcop line, see below), this choice generated peculiar artefacts in the residuals, which change significantly in shape depending on the used scale values. In these two cases, hence, we preferred the use of \textsc{'hogbom'}. The cubes were imaged at the original channel width. Figure~\ref{fig:CO_chmap} shows the channel maps of the five central channels around the local-standard-of-rest velocity of the source for CO $2-1$, similarly to Fig.~\ref{fig:H2CO}.

The \dcop $3-2$ line exhibits very extended emission, and the high spectral resolution of this SPW allows us to identify clear signs of missing flux due to large-scale filtering of the interferometer. We hence combined the interferometric data with an APEX single-dish map of the same transition. These data will be presented in a separate paper (Majerczyk et al., in prep.), but we briefly summarise their key properties here. The data were collected under project ID M9502C\_115. They consist of one on-the-fly datacube with a $2'.5 \times 2'.5$ footprint, enough to encompass the ALMA FoV completely. The data have a sensitivity of $50\, \rm mK$ and a spectral resolution of $61.5 \,$kHz. 
\FloatBarrier
\begin{figure}[!h]
\centering
    \includegraphics[width=\linewidth]{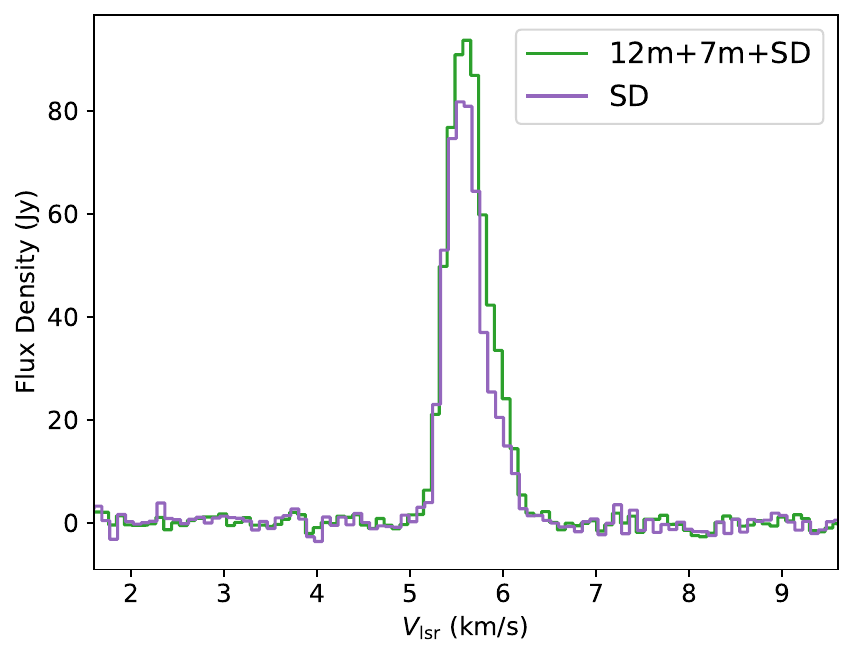}
    \caption{Comparison between the APEX-only spectrum extracted from the central pixel (in purple) and the ALMA+APEX spectrum extracted from a circular region equivalent to the APEX beam size ($30''.6$). The plot shows how the data combination 12m+7m+APEX recovers well the flux seen in the single-dish data. \label{fig:dcop_ALMA_APEX} }
\end{figure}
We performed the data combination using \textsc{sdintimaging}\footnote{We followed the steps presented here: \url{https://casaguides.nrao.edu/index.php/M100_Band3_Combine_6.6.6}.}. As a preparation, we regridded the APEX data to the same spatial and spectral axes as those of the 7m+12m data. We converted the flux scale in $\rm Jy \, beam^{-1}$ using the value $S = 36 \rm \, Jy \, K^{-1}$ found in the APEX available lists of efficiencies\footnote{Available at \url{https://www.apex-telescope.org/telescope/efficiency/?yearBy=2025}.}. The cleaning was performed interactively. We set the parameter \textsc{sdgain=2}, to overweight the single-dish data since they are more sensitive than the ALMA ones in the overlapping UV-space distances. Figure~\ref{fig:dcop_ALMA_APEX} shows the flux recovery obtained with the data combination. 

\FloatBarrier

\section{ SiO position-velocity plot}
{We produced a PV plot starting from the SiO datacube, using the line shown in the right panel of Fig.~\ref{fig:cont}, which follows the axis of the red-shifted and blue-shifted outflow lobes. The result is shown in Fig.~\ref{fig:PV_plot}. }
\FloatBarrier
\begin{figure}[!h]
\centering
    \includegraphics[width=\linewidth]{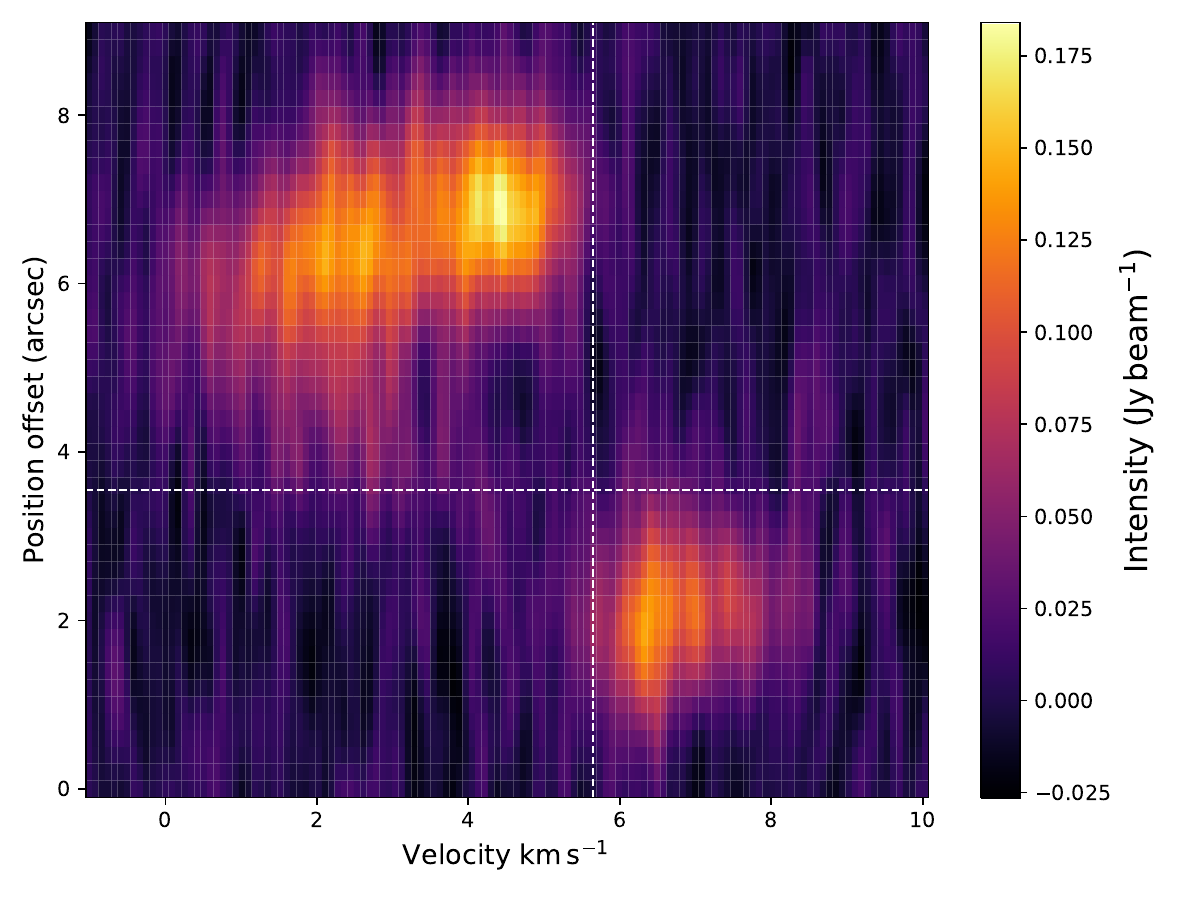}
    \caption{Position-velocity plot obtained from the 7m+12m combined SiO $5-4$ datacube, computed along the axis of the blue-shifted and red-shifted outflow lobes. The vertical and horizontal dashed lines show the position of the central object. \label{fig:PV_plot},}
\end{figure}

\section{The \dcop spectra and modelling results\label{app:dcop}}
\begin{figure*}[hbt]
    \centering
    \includegraphics[width=.63\linewidth]{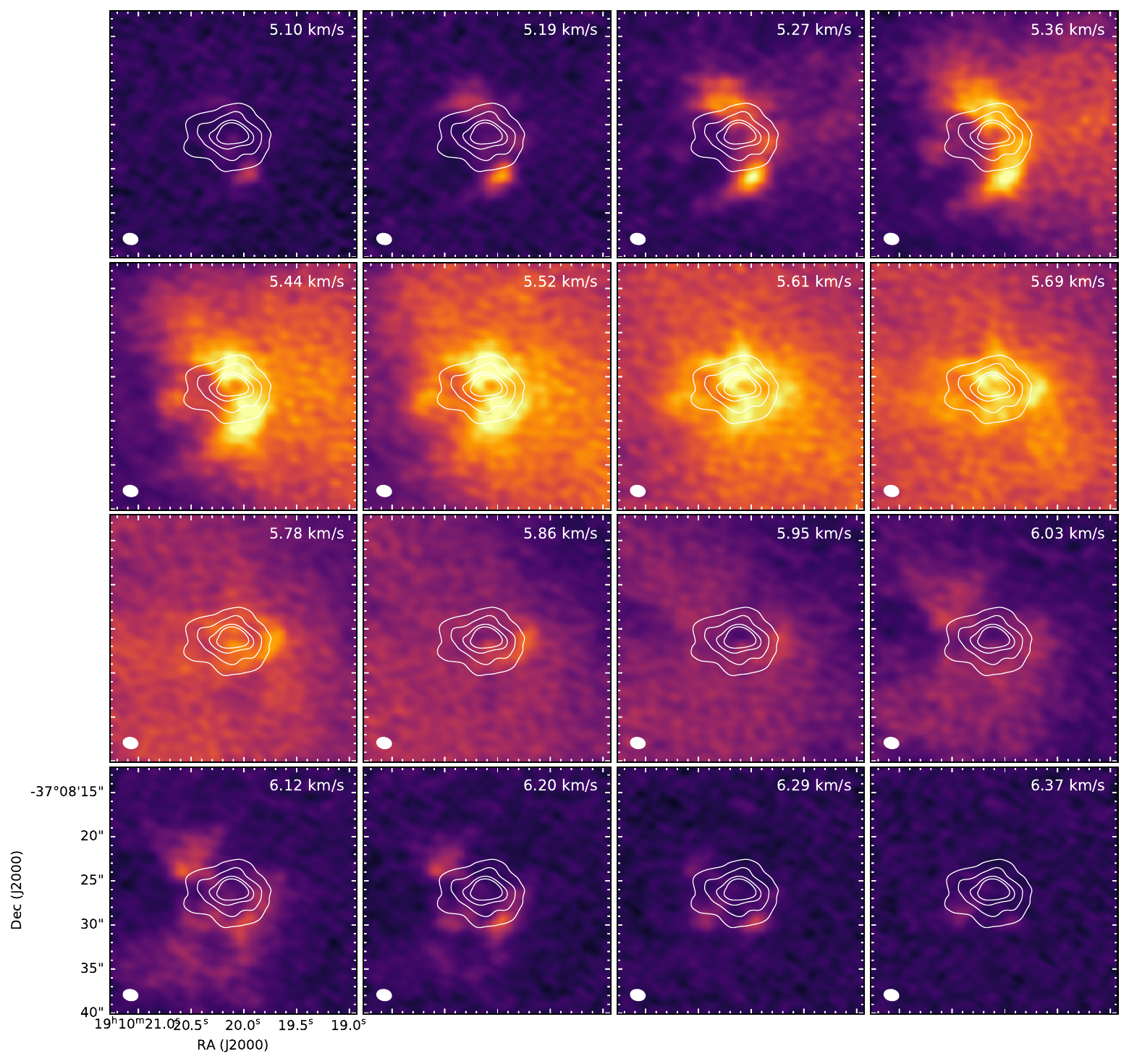}
    \caption{Channel maps of the \dcop $3-2$ transitions. The colorscale is linear, stretching from 0 to $0.5\, \rm Jy \, beam^{-1}$. The velocity of each channel is reported in the top-right corner. The white contours show the continuum emission, as in previous images. The beam size is shown in the bottom-left corner. \label{fig:dcop_channelmap}}
\end{figure*}
We show the channel map obtained from the central 16 channels around the source's  $V_\mathrm{lsr}$ in Fig.~\ref{fig:dcop_channelmap}. The emission spans the velocity range $5.1-6.3\,$\kms. The white contours represent the continuum emission. The \dcop transition at blue velocities appears localised in two main spots north and south of the continuum-traced envelope. Then, the emission spreads over the whole field-of-view, showing a clear gradient moving from the north-east towards the south-west. The areas that emit above $\sim 6\,$\kms are those associated with the second velocity component.

Figure~\ref{fig:dcop_spectra} show a collection of 25 \dcop spectra taken at the position shown in Fig.~\ref{fig:dcop_vel} and Fig.~\ref{fig:dcop_sigma}. The figure is meant to showcase a few examples of positions where just one velocity component is seen, and positions where instead two components are visible and can be fit simultaneously. The best-fit results for the centroid velocity and velocity dispersion of the secondary (red-shifted and weaker) velocity component are shown in Fig.~\ref{fig:dcop_2gauss}. Figure~\ref{fig:dcop_sigma} presents the same plot as of Fig.~\ref{fig:dcop_vel}, but this time for the velocity  dispersion $\sigma_\mathrm{V}$.

 \begin{figure*}[!h]
   \centering
   \includegraphics[width=.63\hsize]{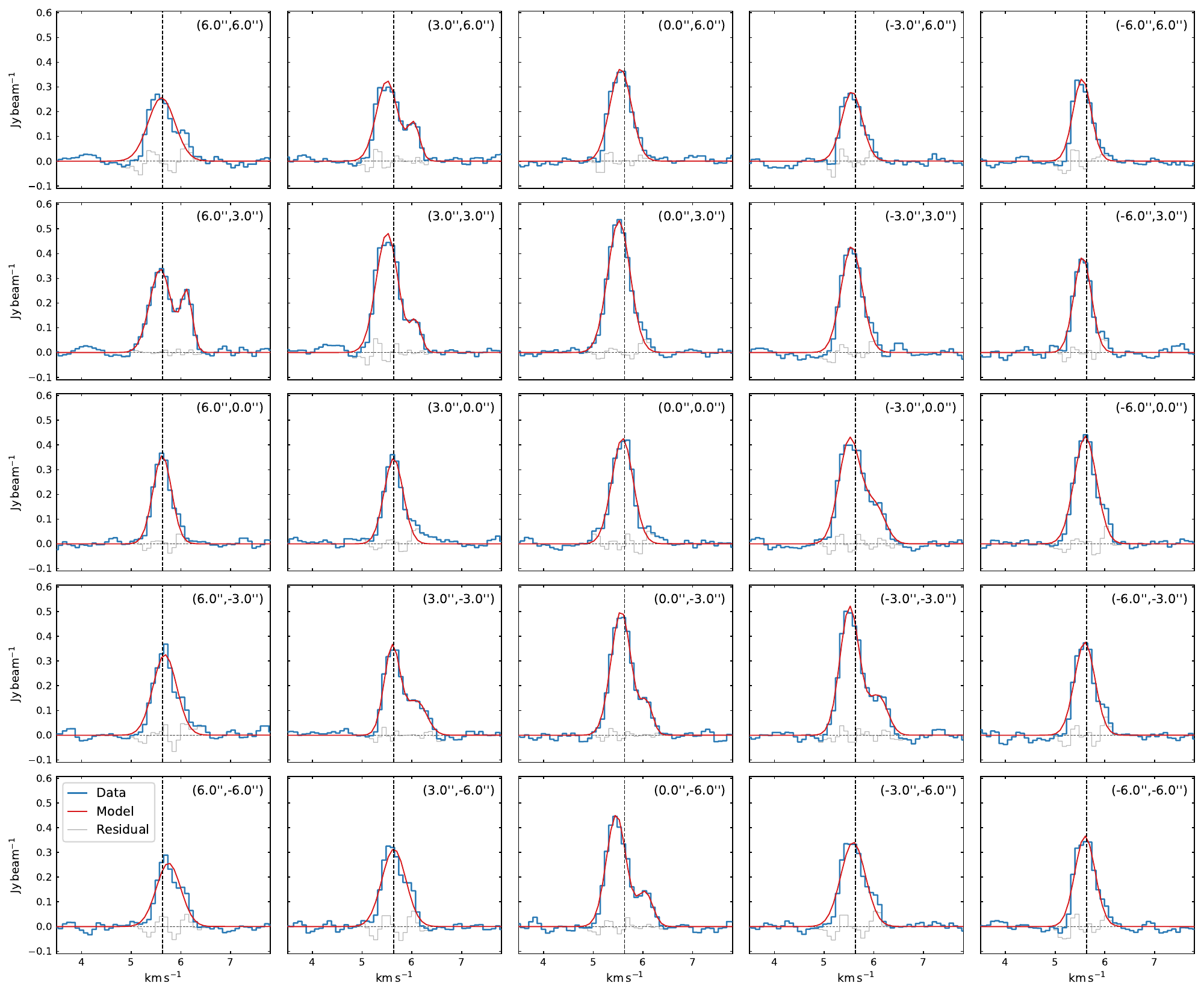}
      \caption{Collection of 25 \dcop $3-2$ spectra, taken at the offsets written in the top-right corners and shown with black crosses in Fig.~\ref{fig:dcop_sigma}. The observed spectra (interferometer and single-dish combined, and primary-beam corrected) are shown with the blue histograms. The best-fit model obtained with either one or two velocity components is shown in red. The residuals are shown in light grey. The vertical dashed line in each panel shows the source velocity $5.65\,$\kms, inferred from APEX data of \nnhp and \nndp \citep{Redaelli25}. Note that, when two components are fit, the blue-shifted component is closer to the source velocity. }
         \label{fig:dcop_spectra}
   \end{figure*}

\begin{figure}[!h]
    \centering
    \includegraphics[width=\linewidth]{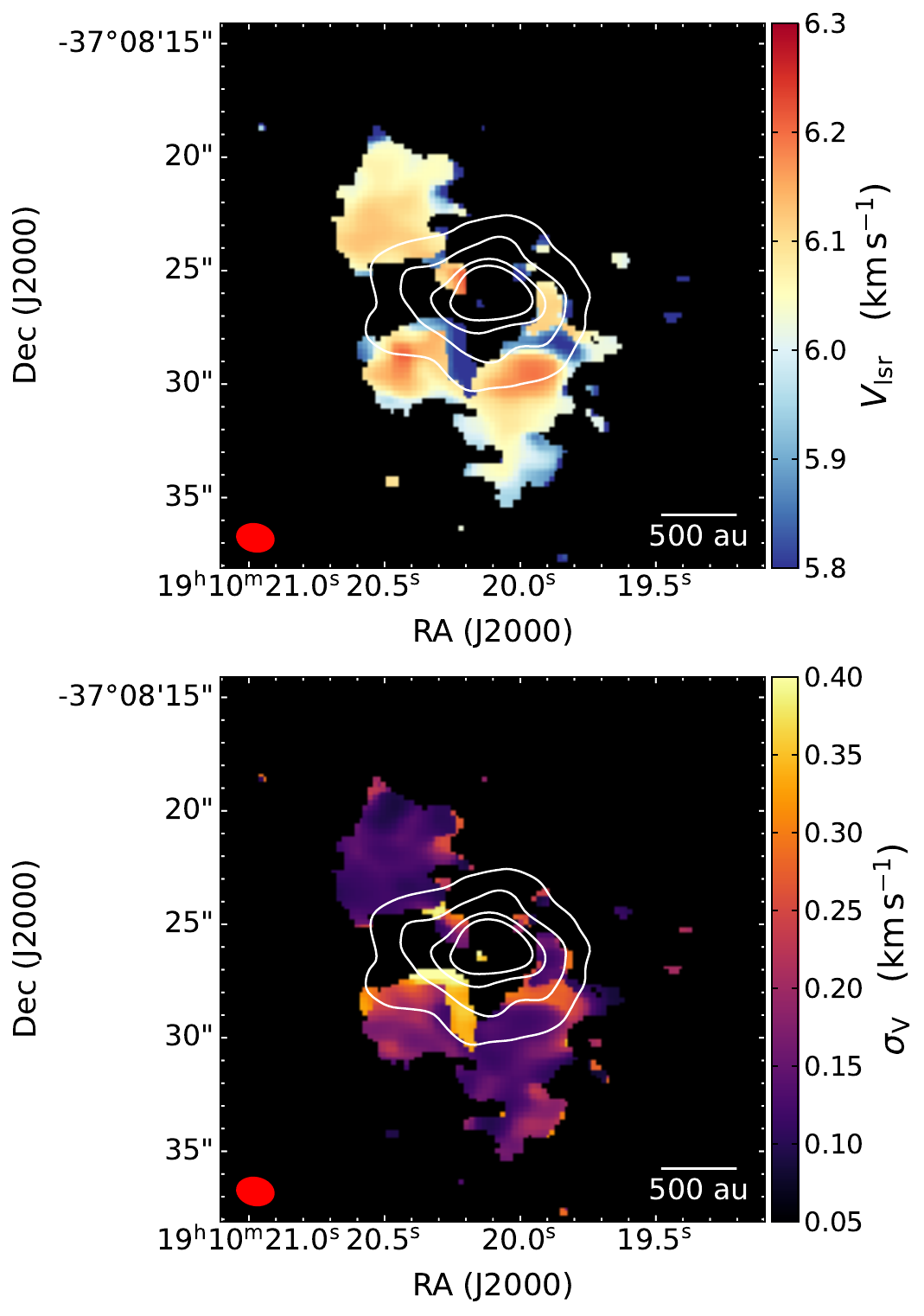}
    \caption{Best-fit solutions for the centroid velocity (left panel) and velocity dispersion (right panel) of the second spectral component at higher velocities. The white contours show the continuum emission, as in previous images. The beam size is shown in the bottom-left corner, and the scale bar in the bottom-right one. }
    \label{fig:dcop_2gauss}
\end{figure}
\begin{figure}[!h]
   \centering
   \includegraphics[width=\linewidth]{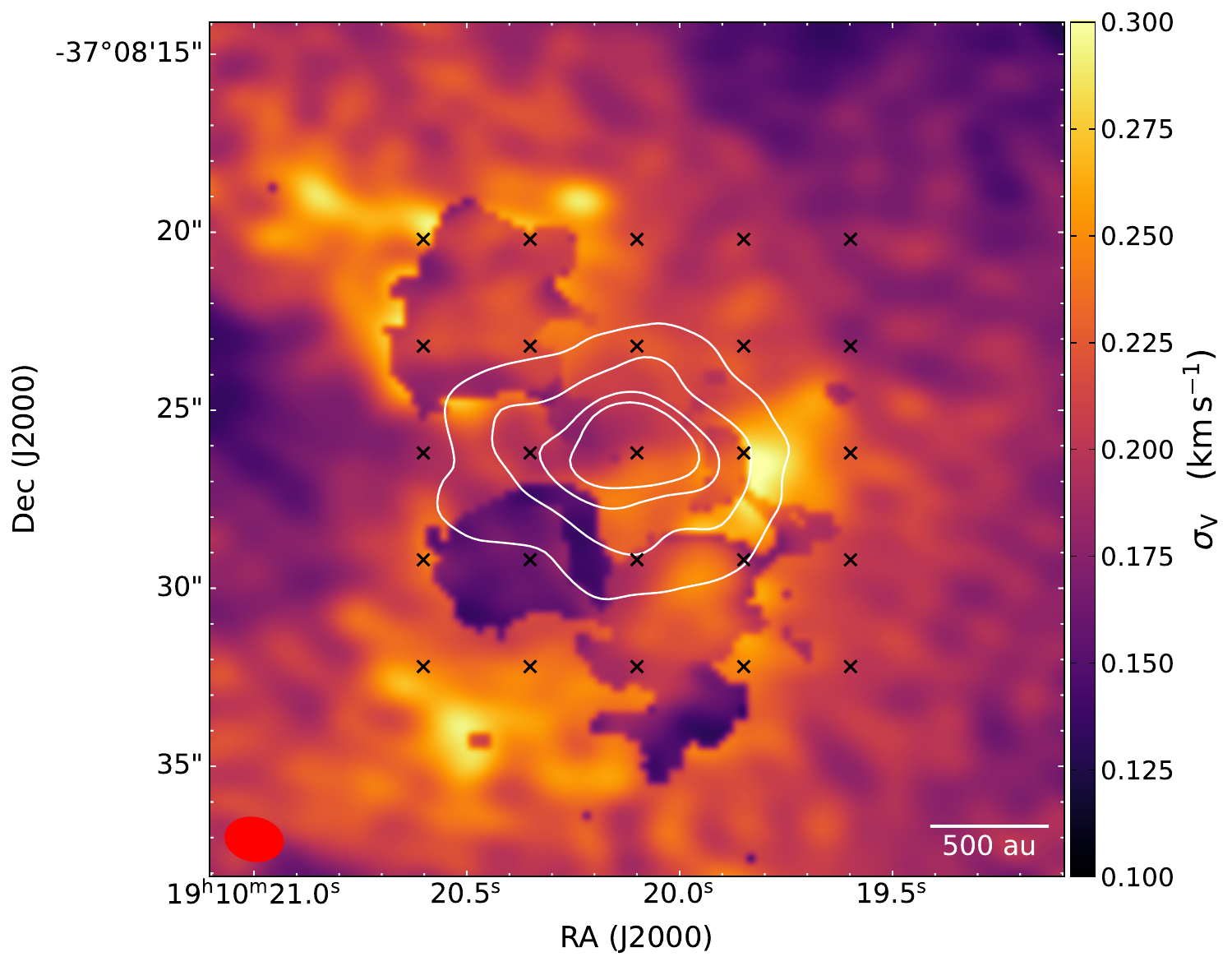}
      \caption{Velocity dispersion obtained by fitting one or two Gaussian component(s) to the \dcop $3-2$ line, as described in the main text, similarly to Fig.~\ref{fig:dcop_vel}. The black crosses show the positions where the spectra in Fig.~\ref{fig:dcop_spectra} are taken.}
         \label{fig:dcop_sigma}
   \end{figure}






\end{appendix}
\end{document}